\documentclass{emulateapj}

\newcommand{\M}{{\cal M}}

\newcommand{\dd}{{\rm d}}

\begin{document}

\title{A simple toy model of the advective-acoustic instability. II.
Numerical simulations}

\author{Jun'ichi Sato, Thierry Foglizzo \& S\'ebastien Fromang $^{1,2}$}
\affil {$^{1}$ CEA, Irfu, SAp, Centre de Saclay, F-91191 Gif-sur-Yvette, France.\\
$^{2}$ UMR AIM, CEA-CNRS-Univ. Paris VII, Centre de Saclay, F-91191 Gif-sur-Yvette, France.
}


\begin{abstract}
The physical processes involved in the advective-acoustic instability
are investigated with 2D numerical simulations. Simple toy models,
developed in a companion paper, are used to describe the coupling
between acoustic and entropy/vorticity waves, produced either by a
stationary shock or by the deceleration of the flow. 
Using two Eulerian codes based on different second order upwind
schemes, we confirm the results of the perturbative analysis. The
numerical convergence with respect to the computation mesh size is
studied with 1D simulations. We demonstrate that the numerical
accuracy of the quantities that depend on the physics of the shock is
limited to a linear convergence. We argue that this property
is likely to be true for most current numerical schemes dealing with
SASI in the core-collapse problem, and could be solved by the use of
advanced techniques for the numerical treatment of the shock. We
propose a strategy to choose the mesh size for an accurate treatment
of the advective-acoustic coupling in future numerical simulations. 
\end{abstract}


\section{Introduction}

Most of our knowledge about the possible consequences of SASI 
on the core-collapse problem has been built, 
over the last 5 years, on the results 
of multidimensional numerical simulations 
\citep[e.g.][]{Blondin+03,Scheck+04,Burrows+06,BM07,MJ07,Iwakami+08}.  
Whether or not SASI can contribute to overcome the explosion threshold, 
to kick the neutron star and alter its spin is still debated. 
In addition to the fundamental uncertainties associated 
with the equation of state of dense matter 
or the numerical treatment  of neutrino transport, 
some difficulties are simply related to multidimensional hydrodynamics 
\citep{Blondin+03, Ohnishi+06, BM06, BM07, Iwakami+08}. 
This latter difficulty is partly due to the complexity 
of the mechanism underlying SASI, 
which is at best unfamiliar, and possibly also affected by the
different numerical 
techniques used by different groups.
The present study aims at improving our understanding 
of the instability mechanism at work 
by studying the advective-acoustic instability in the highly simplified set up 
introduced in the first paper of this series \citep[][~hereafter paper I]{F08}. 
We note that a debate exists about the nature of this mechanism, as
witnessed by \citet[][~hereafter BM06]{BM06},~\citet[][~hereafter FGS07]{Foglizzo+07}, \citet{Laming07}, \citet{YF08}
and \citet{Laming08}. Thus we believe that a better understanding of the
advective-acoustic instability 
in simple examples can help recognise it in more complex situations.
The separation of the advective-acoustic cycle into two separate problems 
is necessary in order to identify, between advected and acoustic perturbations, 
the consequences of each one on the other, as seen on Fig.~7 of
\citet{Blondin+03} or Figs.~11-12 of \citet{Scheck+08}. 
In paper I, the following questions were answered through a
perturbative analysis: 
\par (i) what are the amplitudes of the entropy and vorticity waves generated 
by a shock perturbed by an acoustic wave propagating against the flow, 
towards the shock?
\par (ii) what is the amplitude of the acoustic wave generated 
by the deceleration of an entropy/vorticity wave 
through a localised gravitational potential? 

The first purpose of our study is thus to check the results 
of the perturbative analysis presented in paper I through numerical
experiments, thus providing concrete examples of the coupling
processes involved. 

 The second purpose of this study is to gain confidence in the
  results of more elaborate numerical simulations by assessing their
  accuracy using our simple set up. The 2D numerical simulations of
BM06 showed some 
globally good agreement with the perturbative analysis 
of FGSJ07. 
The typical error on the growth rate and the oscillation frequency of SASI, 
around $30\%$, was not small though. Could this be a concern 
for the many other simulations which use a coarser mesh size? 
We wish to evaluate quantitatively, using our simple toy model, 
to what extent the advective-acoustic instability can be affected 
by numerical resolution. 

The paper is organised as follows. In Sect.~2, 
the set up of the simulations is described and the numerical codes 
are presented. Sect.~3 illustrates qualitatively the two coupling processes 
involved in the advective-acoustic instability using 2D simulations. 
A quantitative analysis of these simulations is also performed which validates 
both the perturbative analysis and the numerical technique. 
In Sect.~4, we evaluate the rate of numerical convergence with respect to 
the mesh size, using a series of 1D numerical simulations. 
While the accuracy of the acoustic feedback produced by the flow gradients 
is quadratic with respect to the mesh size, 
the accuracy of the entropy wave produced by the shock depends 
on the mesh size only linearly. The linear phase of the full problem is simulated in Sect.~5, where the oscillation frequency and growth rate are compared to the results of the perturbative analysis.
The consequences of these numerical difficulties 
for the simulations of core-collapse supernovae are discussed in Sect.~6. 

\section{Numerical techniques and set up of the simulations}

\subsection{Numerical techniques}

The governing equations are solved using the AUSMDV scheme \citep{WL94}, 
which is a second-order finite volume scheme. The former version of
AUSMDV was called 
``advection upstream splitting method'' (AUSM) and developed by  
\citet{LS93}. 
AUSM is a remarkably simple upwind flux vector splitting 
scheme that treats the convective and pressure terms of the flux function 
separately. 
In the AUSMDV, a blending form of AUSM and flux difference is used, and 
the robustness of AUSM in dealing with strong shocks
is improved. 
A great advantage of this scheme is the reduction of numerical viscosity,
which gives sharp preservation of fluid interfaces and 
high resolution feature as in the ``piecewise parabolic method'' 
(PPM) of \citet{CW84}. 
Some advantages over PPM are simplicity and a lower computational cost. 
In Sect.~4, the numerical results obtained with AUSMDV are
  compared with those computed using RAMSES \citep{T02}. RAMSES is
  also a second order shock--capturing code. It
  uses the MUSCL--Hancok scheme to update the MHD equation. For the
  simulations presented in sect.~5, we used the MinMod slope
  limiter along with the HLLD Riemann solver \citep{MK05}, 
  which reduces to the HLLC Riemann solver \citep{T+94} in the
  hydrodynamic case dealt with in this paper.

\subsection{General set up}
\begin{figure}
\plotone{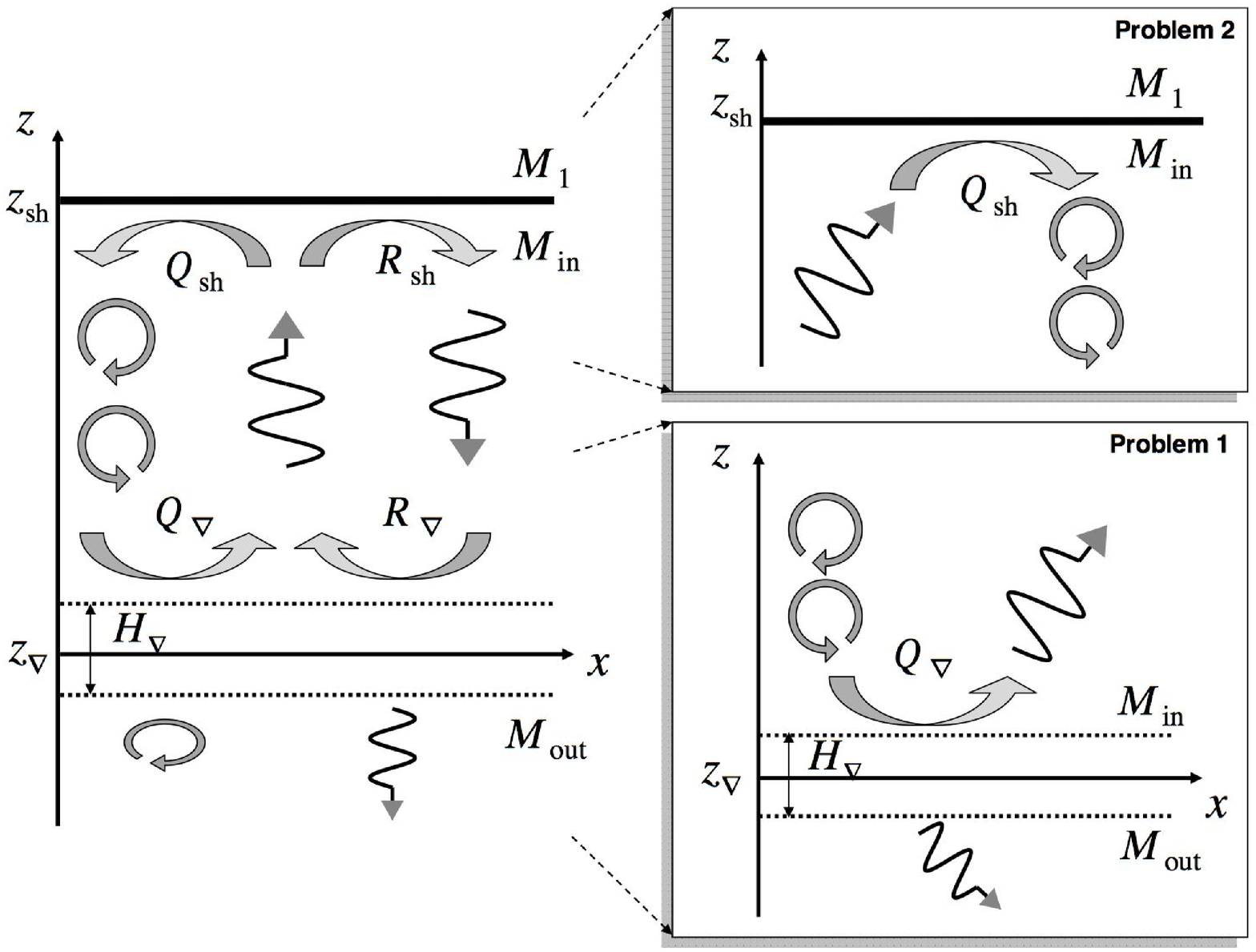}
\caption[dummy]{Schematic view of the advective-acoustic cycle occurring 
in the toy model, separated in two sub problems. 
Entropy/vorticity perturbations are noted as circular arrows, 
while acoustic waves are noted as wavy arrows. The linear coupling between waves is measured by the efficiencies ${\cal Q}_{\rm sh}$, ${\cal Q}_\nabla$, ${\cal R}_{\rm sh}$ and ${\cal R}_\nabla$.}
\label{fig_schematic}
\end{figure}

In this section, we describe the problems we designed to
  illustrate the physical mechanisms underlying the advective--acoustic
  instability. Our ``Problem~1" is aimed at studying the
interaction of waves 
in a stationary subsonic flow decelerated 
across a localised external potential, whereas 
``Problem~2" studies the interaction of waves with a stationary shock 
in a uniform potential. Both problems were described in detail 
in the linear approximation in paper I, and are schematically illustrated 
by Fig.~\ref{fig_schematic}. 
Let us recall that the stationary flow is uniform in the $x$ direction, 
and flows along the $z$ direction with a negative velocity. 
The ideal gas satisfies a polytropic equation of state with an adiabatic index 
$\gamma=4/3$, and a measure of the entropy is defined 
as $S\equiv  (\log(p/\rho^\gamma))/(\gamma-1)$. 
The horizontal size of the computation domain is noted $L_x$. 
The index ``1" refers to the supersonic flow ahead 
of the shock ($z>z_{\rm sh}$), and ``in" refers to the subsonic region 
after the shock ($z<z_{\rm sh}$). ${\cal M}_{\rm in}$, $v_{1}$ 
and $\rho_{1}$ are determined by the Rankine-Hugoniot relations as follows:
\begin{eqnarray}
{\cal M}_{\rm in}&=&\left(
\frac{2+\left(\gamma-1\right){\cal M}_{1}^{2}}
{2\gamma {\cal M}_{1}^{2}-\gamma+1}
\right)^{\frac{1}{2}},\label{RHM}\\
{v_{1}\over v_{\rm in}}&=&
\frac{\left(\gamma+1\right){\cal M}_{1}^{2}}
{2+\left(\gamma-1\right){\cal M}_{1}^{2}},\\
{\rho_{1}\over \rho_{\rm in}}&=&\frac{v_{\rm in}}{v_{1}}, 
\end{eqnarray}
where $v_{\rm in}=-{\cal M}_{\rm in}c_{\rm in}$. 
The incident Mach number is chosen as ${\cal M}_{1}=5$. 
Thus $\M_{\rm in} \sim 0.39$. 

A region of deceleration extends over a width $\sim H_\nabla$ centred 
on $z_\nabla=0$, separating two uniform subsonic regions indexed 
by ``in'' and ``out'', respectively. 
The external potential $\Delta \Phi (z)$ responsible 
for the flow gradients is defined by
\begin{equation}
\Phi(z)\equiv\frac{\Delta \Phi}{2}
\left[
\tanh \left(\frac{z-z_{\nabla}}{H_{\nabla}/2}\right)+1
\right].
\end{equation}
The potential jump $\Delta \Phi>0$ 
is set by specifying the sound speed ratio 
$c_{\rm{in}}/c_{\rm{out}}$: 
\begin{equation}
\Delta \Phi=\left(
\frac{{\cal M}_{\rm out}^{2}}{2}+\frac{1}{\gamma-1}
\right)c_{\rm out}^{2}
-\left(
\frac{{\cal M}_{\rm in}^{2}}{2}+\frac{1}{\gamma-1}
\right)c_{\rm in}^{2}. 
\end{equation}
Defining $H\equiv z_{\rm sh}-z_\nabla$, we adopt $H_{\nabla}/H=0.1$ 
and $c_{\rm in}^2/c_{\rm out}^2=0.75$ in this study, as in paper~I. 

Time is normalised by $\tau_{\rm aac}$, which 
is a reference timescale associated to the advective-acoustic cycle defined 
as follow:
\begin{equation}
\tau_{\rm aac}\equiv
{1\over1-{\cal M}_{\rm in}}\;\;{H\over |v_{\rm in}|}.
\end{equation}
The advection time through the deceleration region $\tau_\nabla$ 
is associated in paper I to a frequency cut-off $\omega_\nabla$, 
above which the efficiency of acoustic feedback decreases:
\begin{eqnarray}
\tau_\nabla&\equiv&\int_{z_{\rm \nabla}-H_\nabla/2}^{z_{\rm \nabla}+H_\nabla/2}{\dd r\over |v|},\\
\omega_\nabla&\sim&{1\over\tau_\nabla}.
\end{eqnarray}
Units are chosen such that $c_{\rm in}=1$, $\rho_{\rm in}=1$
and $H=1$.
Since $p=\rho c^2/\gamma$ and $\gamma=4/3$, then $p_{\rm in}=0.75$ 
and $S_{\rm in}\sim -0.86$. 
The reference timescale is thus $\tau_{\rm aac}\sim 4.2$, 
and $\tau_\nabla\sim 0.41$, 
so that $\omega_\nabla\tau_{\rm aac}/2\pi\sim1.6$. 

Periodic boundary conditions are applied in the $x$-direction. 
Linear perturbations are characterised 
by their wavenumber $k_x\equiv 2\pi n_x/L_x$, with $L_x=4$, and their frequency $\omega_0$. 
With this set of parameters, we expect from paper I a dominant mode $n_x=1$ 
with a growth rate $\omega_i\tau_{\rm aac}=0.22$ 
and an oscillation frequency $\omega_r\tau_{\rm aac}/2\pi=1.13$.\\
With these parameters, the frequency $\omega_{\rm ev}$ below which acoustic waves are evanescent in the $z$ direction is $\omega_{\rm ev}^{\rm in}\tau_{\rm aac}/2\pi=0.96$ in the uniform subsonic region before deceleration, and $\omega_{\rm ev}^{\rm out}\tau_{\rm aac}/2\pi=1.20$ after deceleration (Eq.~(13) in paper~I). For $\omega_r\tau_{\rm aac}/2\pi=1.13$, acoustic waves are evanescent after the region of deceleration with an evanescence length $\lambda_z\sim 1.9H$, deduced from Eq.~(19) in paper~I. 

\subsection{Set up of ``Problem 1"}
 
In ``Problem~1", the flow is only composed of three parts, without a
shock, and is thus entirely subsonic. 
Once the stationary unperturbed flow is well established on the
computation grid, an entropy/vorticity wave is generated at the upper
boundary, at $z=3$. This wave is in pressure equilibrium ($\delta
p=0$). The corresponding perturbations of entropy $\delta S$ and
density $\delta \rho$ are defined as follows: 
\begin{eqnarray}
\delta S &\equiv& \epsilon_S \cos \left(-\omega_{0}t+k_{x}x+k_{z}z\right),\\
{\delta \rho\over\rho_{\rm in}}  &\equiv&
\exp\left(-\frac{\gamma-1}{\gamma}\delta S\right)-1\sim -\frac{\gamma-1}{\gamma}\delta S.
\end{eqnarray}
where $\epsilon_S=10^{-3}$ is the parameter defining the amplitude of the
entropy perturbation. The vertical wavenumber of an advected wave is
$k_{z}=\omega_{0}/v_{\rm in}$.  
The incompressible velocity perturbations $\delta v_{x}$ are $\delta
v_{z}$ are chosen such that the vorticity $\delta w_y$ is the same as
when produced by a shock (Eqs.~(A6-A9) in paper I): 
\begin{eqnarray}
\delta v_{x} &\equiv& {k_{x}\omega_{0}c_{\rm{in}}^{2}\over\omega_{0}^{2}+k_{x}^{2}v_{\rm in}^{2}}
\;\;{\delta S\over\gamma} ,\\
\delta v_{z} &\equiv&-{k_x^2v_{\rm in}c_{\rm in}^{2}\over\omega_{0}^{2}+k_{x}^{2}v_{\rm in}^{2}}
\;\;{\delta S\over\gamma} ,\\
\delta w_y&=&-{k_xc_{\rm in}^2\over v_{\rm in}}\;\;{\epsilon_S\over\gamma} \sin \left(-\omega_{0}t+k_{x}x+k_{z}z\right).
\end{eqnarray}
We choose free boundary conditions at the lower boundary ($z=-5$),
sufficiently far from the shock to avoid any effect from a reflected wave. 
Between $z=-2$ and $-5$, we use an inhomogenous mesh 
whose interval increases gradually in the negative $z$-direction. 
We perform simulations with $k_{x}=2\pi/L_{x}$ and different
values of the frequency $\omega_0$ and mesh size $\Delta z$. 
The results of the simulations are analysed in Sect.~3.1
and 3.2.

\subsection{Set up for ``Problem 2"}

In our ``Problem 2", the unperturbed stationary flow is composed of two
semi-infinite uniform regions separated by a stationary shock. 
Once the steady flow is  well established on the numerical grid, an
acoustic wave is generated at the lower boundary of the computing
box, at $z=-2$ and propagates against the flow towards the shock. 
The density perturbation $\delta \rho$, 
the pressure perturbation $\delta p$ 
and the velocity perturbations $\delta v_{x}$ and $\delta v_{z}$ 
are defined according to paper I as follows at the lower boundary:
\begin{eqnarray}
\frac{\delta \rho}{\rho_{\rm in}}&\equiv&
\frac{1+\mu{\cal M}_{\rm in}}{1-{\cal M}_{\rm in}^{2}}
\times \epsilon_\rho \cos \left(
-\omega_{0}t+k_{x}x+k_{z}^-z
\right),\\
\frac{\delta p}{p_{\rm in}}&\equiv&
\left(1+\frac{\delta \rho}{\rho_{\rm in}}\right)^{\gamma}-1,\\
\delta v_{x}&\equiv&
\frac{k_{x}c_{\rm in}^{2}}{\omega_{0}}
\times \epsilon_\rho \cos \left(
-\omega_{0}t+k_{x}x+k_{z}^-z
\right),\\
\delta v_{z}&\equiv&
\frac{\mu+{\cal M}_{\rm in}}{1-{\cal M}_{\rm in}^{2}}c_{\rm in}
\times \epsilon_\rho \cos \left(
-\omega_{0}t+k_{x}x+k_{z}^-z
\right),
\end{eqnarray}
where 
\begin{equation}
\mu\equiv\left[
1-\frac{k_{x}^{2}c_{\rm in}^{2}}{\omega_{0}^{2}}
\left(1-{\cal M}_{\rm in}^{2}\right)
\right]^{\frac{1}{2}}, 
\end{equation}
Here $\epsilon_\rho=10^{-3}$ sets the amplitude of the density perturbation. The
vertical wavenumber $k_z^-$ for an acoustic perturbation is given by
Eq.~(19) of paper I:
\begin{eqnarray}
k_z^\pm={\omega\over c_{\rm in}}\;\;{\M_{\rm in}\mp\mu\over 1-\M_{\rm in}^2}.
\end{eqnarray}
We choose fixed boundary conditions at the upper boundary ($z=2$).
The results of the simulations are analysed in section 3.3 and 3.4.

\section{Numerical illustration of the coupling processes and comparison with the linear analysis}

\subsection{Acoustic feedback from the deceleration of a vorticity wave 
(Problem 1)}

\begin{figure}
\plotone{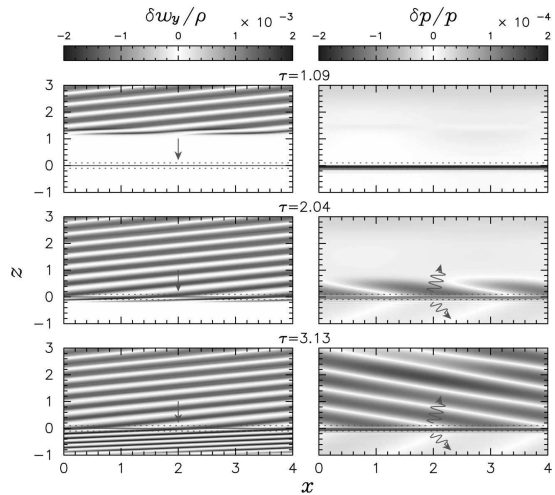}
\caption[dummy]{Production of an acoustic wave by the deceleration of a vorticity wave (Problem 1).
The specific vorticity $\delta w_y/\rho$ (left) and the normalised  pressure perturbation $\delta p/p$ (right) are shown at three successive times, before and after the advected wave reaches the deceleration region localised around $z=0$ (within the dashed lines). 
The parameters are $\omega_0\tau_{\rm aac}/2\pi=2$, 
and $\Delta x=\Delta z=10^{-2}$.
}
\label{fig_P1}
\end{figure}
The snapshots in Fig.~\ref{fig_P1} show the specific vorticity 
$\delta w_y/\rho$ (left column) and pressure perturbation $\delta p/p$ 
(right column) in the flow at three successive times, before and after the moment when the advected wave reaches the deceleration region. The right column of Fig.~\ref{fig_P1} demonstrates the
  absence of an acoustic perturbation until the advected wave reaches
  the region of deceleration. Two acoustic waves are then generated,
  propagating upward and downward. This simple experiment gives a
  concrete illustration of the physical process described in
  analytical terms in paper I. In the bottom plots of
  Fig.~\ref{fig_P1}, the flow has reached the asymptotic regime
  described by a single frequency in paper~I, in which a more quantitative
  comparison of coupling efficiencies can be made. Since the
  computation domain is finite, 
the numerical experiment is stopped before the acoustic waves reach
the vertical boundaries of the computation box in order to avoid
spurious reflections.  The time needed to reach the asymptotic regime described by a single frequency in paper~I depends strongly on the frequency of the wave, and can become prohibitively long close to the frequency of horizontal propagation $\omega_{\rm ev}^{\rm in}$. This can be understood by viewing the semi-infinite acoustic plane wave, involved in both Problems 1 and 2, as an infinite plane wave of frequency $\omega_0$, multiplied by a step function, whose Fourier transform involves a continuum of frequencies. In 1-D, all frequencies would propagate with the same velocity, and the shape of the wave packet would stay unchanged during propagation. In 2D however, the high frequency part of the acoustic spectrum $\omega>\omega_0$ propagates more vertically than the main component, while the low frequency part $\omega_0>\omega>\omega_{\rm ev}^{\rm in}$ propagates more horizontally: this dispersion requires a longer numerical simulation, and thus a larger computational domain in order to avoid acoustic reflections. For this reason we have limited our investigation to the frequencies $\omega_0\tau_{\rm aac}/2\pi=1.5$, $2$, $4$, and $6$.
Note that if the frequency of the perturbation had been chosen 
below the threshold of acoustic propagation 
($\omega<\omega_{\rm ev}^{\rm in}$), 
the acoustic feedback would be evanescent above the deceleration region 
(paper I and Guilet, Sato \& Foglizzo, in preparation).

\subsection{Measure of the acoustic feedback in Problem 1}

\begin{figure}
\plotone{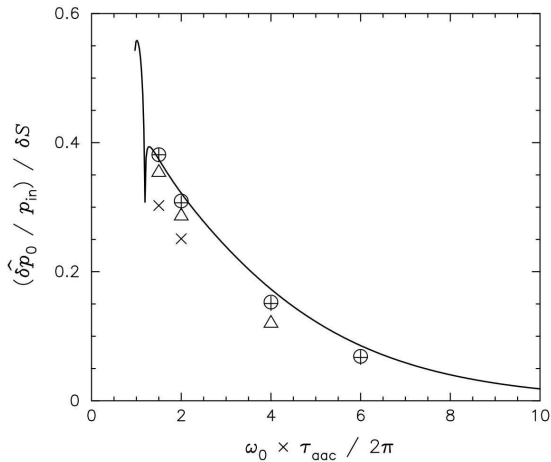}
\caption[dummy]{
Efficiency $(\hat{\delta p}_0 / p_{\rm{in}}) / \delta S$ of the production 
of acoustic waves by the deceleration of entropy/vorticity waves, 
measured at $z=0.5$, as a function of $\omega_{0}$ in Problem~1. 
The solid line shows the curve computed by a linear analysis (paper I). 
The results of numerical simulations are shown for different square mesh sizes
$\Delta x=\Delta z=5\times10^{-2}$ (crosses), $2\times10^{-2}$ (triangles) 
and $10^{-2}$ (circles). The results for $\Delta x=2\times10^{-2}$, $\Delta z=10^{-2}$ are also shown (pluses).
}
\label{omg_dppdS}
\end{figure}

The amplitude of the acoustic feedback is measured in the numerical experiment by using a Fourier transform, in time, of the pressure perturbation over the period $T\equiv 2\pi/\omega_{0}$ of the wave:
\begin{equation}
\hat{\delta p}_0=
\frac{2}{T}\int^{T}_{0}{\delta p}\;
e^{i \omega_{0}t}dt, 
\end{equation}
The symbols in Fig.~\ref{omg_dppdS} are measured 
at $z=0.5$, in a region where the gravitational potential is uniform. 
The full line in Fig. \ref{omg_dppdS} shows the expected efficiency 
$(\hat{\delta p}_0 / p_{\rm in}) / \delta S$ of the acoustic feedback obtained 
by integrating the differential system as in paper I. 
The good agreement with the perturbative calculation for a fine mesh (circles) 
confirms the validity of both the perturbative calculation 
and the numerical code. 
Given the long horizontal wavelength of the perturbations, the results are insensitive to an increase of the horizontal size $\Delta x$ of the mesh (pluses and circles).
As described in paper I, the efficiency of the acoustic feedback decreases for frequencies above the cut-off $\omega_\nabla\sim 1/\tau_{\nabla}$.

\subsection{Entropy/vorticity produced by a shock perturbed 
by an acoustic wave (Problem 2)}

\begin{figure}
\plotone{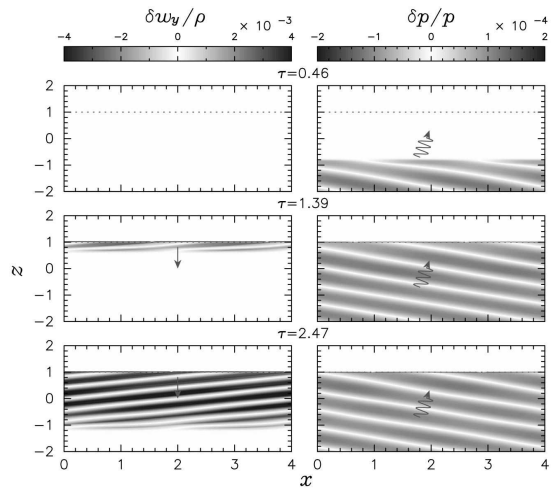}
\caption[dummy]{Production of a vorticity wave by the interaction 
of an oblique acoustic wave with the shock (Problem 2). 
$\delta w_y/\rho$ (left) and $\delta p/p$ (right) are shown 
at three successive times, before and after the acoustic wave reaches 
the shock localised at $z=1$ (dashed line).
A vorticity wave is generated and advected downward.
The parameters are $\omega_0\tau_{\rm aac}/2\pi=2$, 
and $\Delta x=\Delta z=10^{-2}$.
}
\label{fig_P2}
\end{figure}

The upward propagation of the acoustic wave generated at the lower boundary of the computation domain in Problem~2 is visible on the right column of Fig.~\ref{fig_P2}. The three snapshots illustrate the independence of advected and acoustic perturbations in the uniform part of the flow: the vorticity wave visible on the left column in Fig.~\ref{fig_P2} is generated only as the acoustic wave reaches the shock.
This vorticity wave is then continuously generated by the shock and advected downward with the flow. An entropy wave (not shown) is also generated at the shock, with the same appearance as the vorticity wave. 
The lower boundary condition in this experiment is chosen far enough
so that the reflected acoustic wave generated at the shock does not
have time to interact with the lower boundary. The efficiency of
entropy/vorticity generation at the shock can be measured at the time
corresponding to the bottom panel in Fig.~\ref{fig_P2}, and compared
to the calculations of paper I.

\subsection{Measure of the entropy production in Problem 2}

\begin{figure}
\plotone{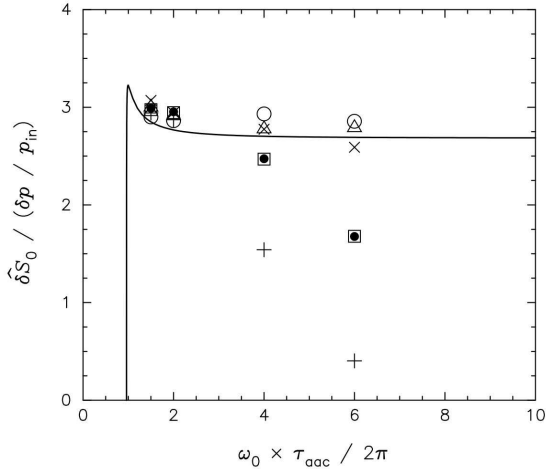}
\caption[dummy]{
Dependence of $\hat{\delta S}_0/(\delta p/p_{\rm in})$, measured at 
$z=0.5$, on the frequency $\omega_{0}$, in Problem~2. 
The solid line shows curve predicted from linear analysis (paper I). 
The result of numerical simulations is shown for different mesh sizes
$\Delta z=2\times10^{-2}$ (pluses), $\Delta z=10^{-2}$ (squares), $5\times10^{-3}$ (crosses), 
$2\times10^{-3}$ (triangles) and $10^{-3}$ 
(circles) where $\Delta x=2\times10^{-2}$.
The filled points show the results for $\Delta x=\Delta z=10^{-2}$. 
}
\label{omg_dS}
\end{figure}

According to Eqs.~(30-31) of paper I, the amplitude 
$\delta S_{\rm th}$ of the entropy wave
produced by an acoustic wave reaching the shock is expected to be related 
to the frequency of the pressure wave as shown by the full line 
in Fig. \ref{omg_dS}:
\begin{eqnarray}
\delta S_{\rm th}=&&
\frac{\delta p}{p_{\rm in}}
\;\;\frac{2}{\cal M_{\rm in}}
\;\;\frac{1-{\cal M}_{\rm in}^{2}}{1+\gamma {\cal M}_{\rm in}^{2}}
\left(1-\frac{{\cal M}_{\rm in}^{2}}{{\cal M}_{1}^{2}}\right)
\nonumber
\\
&& \times
\frac{\mu}{\mu^{2}+2\mu {\cal M}_{\rm in}+{\cal M}_{1}^{-2}}.
\end{eqnarray}
Measuring the amplitude of the entropy wave produced by the shock 
in the numerical simulations is not straightforward 
because of the presence of spurious high frequency oscillations, 
analysed in more details in the next section. 
We choose to measure (at $z=0.5$) 
its fundamental Fourier component $\hat{\delta S}_0$ at the frequency $\omega_0$, 
thus filtering out oscillations at higher frequency. The result
 is displayed in Fig. \ref{omg_dS} for different frequencies and mesh sizes.
We did not notice any dependence on the horizontal size $\Delta x$ of the mesh, for the long horizontal wavelengths considered. The expectation of the perturbative calculation is confirmed, but the convergence to the analytical formula is apparently much slower than for Problem~1. The rate of convergence is analysed in the next section using 1D simulations.

\section{Accuracy of the numerical convergence}

The dependence of the numerical error on the mesh size is easier to
investigate using 1D simulations because of the shorter computation
time. Without excluding the possibility of additional difficulties in
2D, we demonstrate here that some numerical difficulties associated
to the advective-acoustic coupling are already present in 1D. The set
up we use in this section is the same as used for the 2D simulations
except that $k_x=0$.

\subsection{Quadratic convergence in Problem~1}

\begin{figure}
\plotone{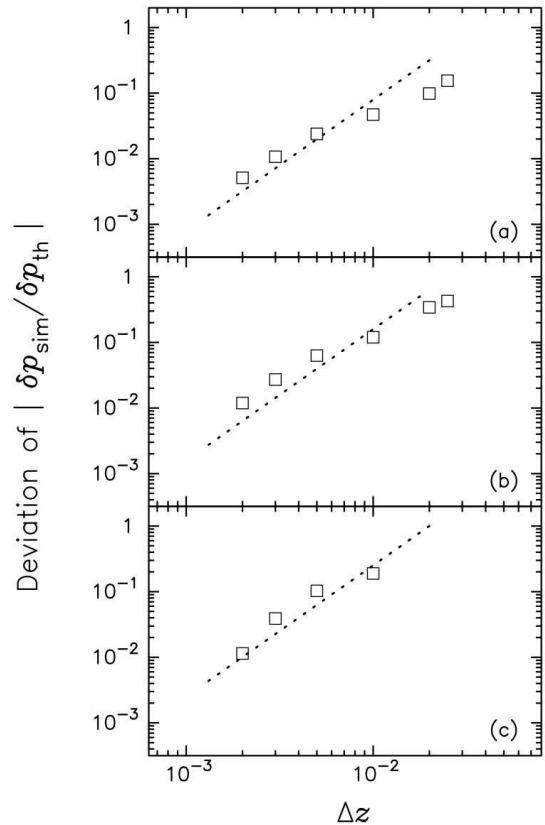}
\caption[dummy]{
Numerical error as a function of the mesh size for Problem~1. 
The panels (a), (b) and (c) correspond to the cases of 
$\omega_{0}\tau_{\rm aac}/2\pi=2$, $4$ and $6$, respectively. 
The dotted lines, proportional to $\Delta z^{2}$, 
illustrate the quadratic convergence. 
}
\label{fig_errP1}
\end{figure}
A series of numerical simulations of Problem~1 in 1D with different
mesh sizes and perturbation frequencies allowed us to measure the
accuracy of the computation compared to the perturbative analysis as
shown in Fig.~\ref{fig_errP1} by the open squares. They are to be
compared with the dotted line, whose slope of $+2$ illustrates
second order convergence for this problem. Remembering that the accuracy of our numerical scheme is second order in space, it is satisfactory to find that the error displayed in Fig.~\ref{fig_errP1} is approximately quadratic with respect to the mesh size. 
The shortest wavelength in Problem~1 is the wavelength $2\pi v_{\rm out}/\omega_0$ of advected perturbations after their deceleration, which is equal to $\sim 0.12$ for the frequency $\omega_0\tau_{\rm aac}/2\pi=6$. We conclude from Fig.~\ref{fig_errP1} that our numerical treatment of advection, propagation and advective-acoustic coupling involved in Problem~1 is accurate at the percent level even when the shortest wavelength is sampled by only $N\sim 10$ grid zones.

\subsection{Linear convergence in Problem 2}

\begin{figure}
\plotone{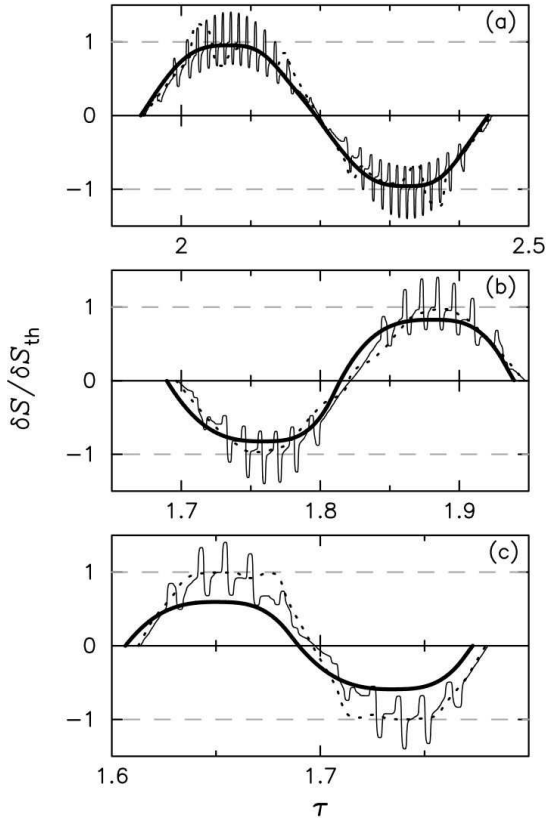}
\caption[dummy]{
Time evolution of the amplitude of $\delta S / \delta S_{th}$ at $z=0.5$ for the same three frequencies 
as in Fig.~\ref{fig_errP1}. 
The thick line, dotted and thin lines correspond to the cases
$\Delta z=10^{-2}$, $10^{-3}$ and $10^{-4}$, respectively. 
}
\label{tau_dS}
\end{figure}

\begin{figure}
\plotone{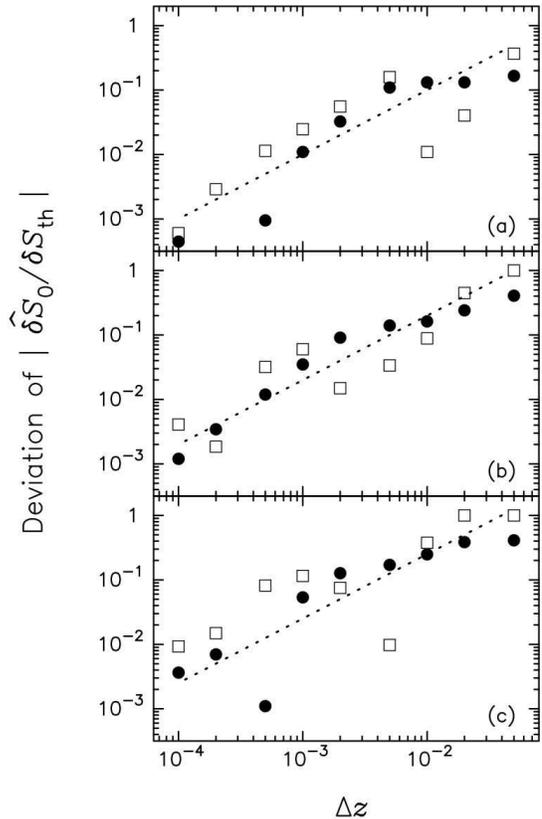}
\caption[dummy]{
Numerical error of the quantity 
$|\hat{\delta S_0}/\delta S_{\rm th}|$ 
as a function of the mesh size for Problem~2. The frequencies 
are the same as in Fig.~\ref{fig_errP1}. 
The empty squares and filled circles were obtained with the AUSMDV scheme and the code RAMSES respectively. The dotted lines, proportional to $\Delta z$, 
illustrate the linear convergence. 
}
\label{fig_errP2}
\end{figure}

Applying the same test to Problem~2 is more complicated because of the high frequency oscillations already mentioned in Sect.~3. The shape of the entropy wave is shown in Fig.~\ref{tau_dS} for different frequencies and mesh sizes. The finer the mesh the higher the frequency of these spurious oscillations. 
We checked that the power involved in the Fourier
  component associated with these higher frequencies is always negligible compared to the main component. The Fourier component associated with the fundamental frequency $\omega_0$ 
  converges slowly to the expected analytical value for a fine mesh.
  The squares in Fig.~\ref{fig_errP2} show the numerical accuracy of the AUSMDV scheme for Problem~2, revealing a linear convergence with the mesh size (as shown by the dotted line of slope $+1$).
We note that a coarse resolution can either underestimate or even
overestimate the production of entropy at the shock. In order to
  show that this linear convergence is not a peculiarity of the AUSMDV
  scheme, these simulations were repeated with the code RAMSES. The
  results obtained with RAMSES are shown by the blacks circles in
  Fig.~\ref{fig_errP2} (note that we also observed spurious
  high frequency oscillations in that case). They are comparable to those obtained using the AUSMDV
  scheme. Based on this comparison, we anticipate that all
  finite volume codes in which the treatment of the shock relies on an
  upwind technique are likely to share the same difficulty:
  quantities produced at the shock location, such as vorticity
  and entropy waves, or the reflected acoustic wave, are
  computed with a first order accuracy with respect to the mesh size. 
  Likewise, we anticipate that all finite volume codes will
  suffer from the presence of spurious high frequency oscillations
  similar to those described above. It is indeed well known that such
  codes are subject to this problem, especially in the case of
  standing shocks, as was reported by \citet{CW84}. In the
  present case, the problem is made worse by the interaction between
  the shock and the sound wave (in the absence of the latter, we
  barely detected high frequency oscillations, with an amplitude of
  the order of $0.5 \%$ of the amplitude of the reflected entropy
  wave). As described by \citet{CW84}, any 
  additional source of dissipation (artificial viscosity, grid
  translation) will result in a decrease of the
  amplitude of the oscillations. For example, with RAMSES, the use of
  the Monotonised Central slope limiter \citep{toro97}, which is known to be less dissipative than MinMod, resulted in the amplitude of the oscillations being
  about three times larger. However, the complete stabilisation of the
  oscillations (through the use of artificial viscosity for example)
  would most probably come at the cost of reducing the growth 
  rate, which we show in Sect.~5 not to be affected by the oscillations.

\section{Eigenfrequency in the full toy model}

\begin{figure}
\plotone{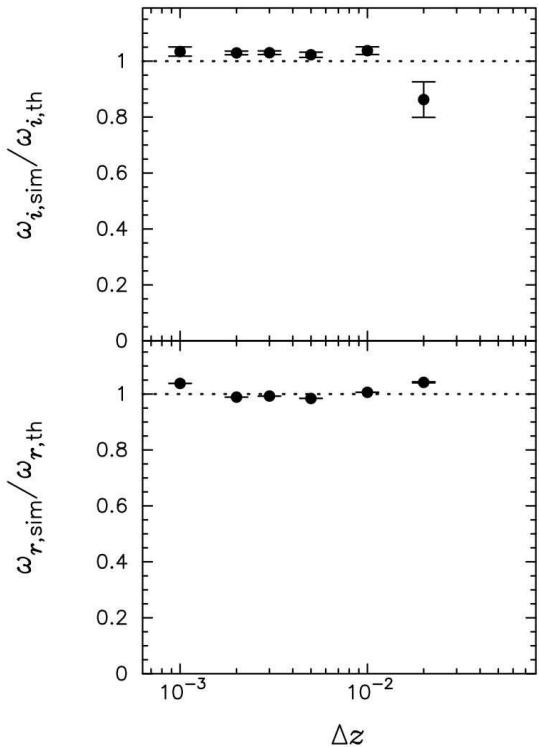}
\caption[dummy]{Growth rate $\omega_{i,\rm sim}$ and oscillation frequency $\omega_{r,\rm sim}$ of the most unstable mode ($n_x=1$) measured in a numerical simulation of the full toy model, compared to the values $\omega_{i,\rm th}$, $\omega_{r,\rm th}$ obtained from the perturbative analysis (paper~I). The parameters are $L_x/H=4$, $H_\nabla/H=0.1$, $\M_1=5$, $c_{\rm in}^2/c_{\rm out}^2=0.75$, and $\Delta x=10^{-2}$. Perturbations were initiated with a random noise. Error bars are associated to the fitting procedure.
}
\label{fig_full}
\end{figure}

The full toy model has been simulated in order to measure the oscillation frequency $\omega_r$ and growth rate $\omega_i$ of the dominant eigenmode for $\M_1=5$, $H_\nabla/H=0.1$, $L_x/H=4$ and $c_{\rm in}^2/c_{\rm out}^2=0.75$. One difficulty for this simulation is the numerical relaxation of the unperturbed flow on the computational grid, which can result in a slow drift of the shock. The stationary flow is constructed by first obtaining a stationary subsonic flow in the gravitational potential, and then choose the upstream flow such that a shock is stationary at $z= z_{\rm sh}$. As a result of numerical discretization, the upstream mach number may slightly differ from $\M_1=5$, by a few percents. This difference is taken into account in the perturbative calculation of the reference eigenfrequency. Perturbations are incorporated as a random noise in the transverse velocity at the level of $10\%$ of the flow velocity in the uniform region between $z=0.3$ and $z=0.9$. The linear evolution is dominated by the mode $n_x=1$, as expected from the linear stability analysis. 
The comparison with the perturbative calculation is shown in Fig.~\ref{fig_full}.
The oscillation frequency and growth rate, determined numerically, are accurate to about $5\%$ for $\Delta z\le10^{-2}$, suggesting that the spurious high frequency oscillations revealed in Sect.~3.4 and 4.2 have a minor effect on the eigenfrequency of the most unstable mode.
The slight excess of the growth rate $\omega_{i,\rm sim}$ in Fig.~\ref{fig_full} may be related to the fact that entropy and vorticity pertrubations are slightly overproduced at the shock, as seen in Fig.~5 for Problem~2. This effect, however, should be partially compensated by the slight underproduction of the acoustic feedback in Problem~1 (Fig.~3).\\
A significant damping of the instability ($\sim 14\%$) occurs if the grid is too coarse $(\Delta z=2\times10^{-2})$ but even then, the oscillation frequency is accurate within $5\%$. The surprising accuracy of the oscillation frequency can be understood by the fact that the oscillation timescale is closely related to the timescale, for an advective-acoustic cycle between the shock and the deceleration region. Since the position of the acoustic feedback is set by the external potential in our toy model, this timescale barely depends on the numerical resolution. One must keep in mind that in a realistic flow where gradients are due to cooling processes, a change of numerical resolution could influence the position of the deceleration region, and could thus affect the oscillation timescale of the instability.  

\section{Consequences for core-collapse simulations}

The results of our numerical experiments can be helpful to choose the mesh size in future simulations of a collapsing stellar core, both at the shock and near the neutron star, in order to make sure that the physics of SASI is correctly treated, at least in the linear regime. Of course, the influence of SASI on the mechanism of core-collapse supernovae depends on non-linear quantities such as the amplitude of the shock oscillations, the advection time through the gain region, or the spectral distribution of energy below the shock. 
Characterising which of the non-linear properties of SASI are most sensitive to the numerical technique is beyond the scope of the present study, and will be investigated in a forthcoming publication. 
We believe however that the coupling between entropy, vorticity and pressure is likely to play an important role even in the non linear regime of SASI, both through the flow gradients and at the shock. The wide range of frequencies involved in the non linear evolution of SASI (e.g. Yoshida et al. 2007) suggests that the accuracy of the numerical treatment should not be limited to the low frequency of the most unstable mode. In this sense, the numerical constraints deduced from our linear analysis should be considered as a minimum requirement, even-though some non-linear consequences of SASI may be less sensitive to numerical resolution than others:  the addition of numerical errors with opposite signs, mentioned in Sect.~5, may contribute to the complex, non monotonic dependence of the explosion time with respect to the numerical resolution, observed by Murphy \& Burrows (2008).

\subsection{Mesh size in the deceleration region}

When the shock stalls above the proto-neutron star, the flow deceleration close to the neutron star is dominated by cooling processes much more than by gravity, and the advective-acoustic coupling there is not adiabatic. By making the choice of simplicity, our toy model does not aim at reproducing quantitatively the efficiency of the acoustic feedback in a non-adiabatic flow. It helps understand that a simulation with a coarse grid in the vicinity of the neutron star may be unable to take into account a possible acoustic feedback from this region, simply because advected perturbations are numerically damped before reaching it. Let us consider a numerical simulation of an advective-acoustic cycle dominated by the oscillation frequency $\omega_0$. The choice of the mesh size close to the surface of the neutron star is not obvious because the wavelength of advected perturbations $\lambda_{\rm adv}\sim 2\pi v(r)/\omega_0$ shrinks as the gas is decelerated. Fortunately, an accurate advection of this perturbation is needed only down to the region where most of the acoustic feedback is generated, adiabatic or not. Since the timescale of the advective-acoustic cycle is larger than the advection timescale, and comparable to the oscillation timescale $2\pi/\omega_{\rm f}$ of the fundamental mode, the region of feedback is necessarily above the radius $r_{\rm in}$ reached 
by the gas during one SASI oscillation. According to Figs.~4 and 5 of FGSJ07, the dominant mode is the fundamental one ($\omega_f=\omega_0$) if the shock is close to the neutron star, or the first harmonic ($\omega_f\sim 2\omega_0$) if the shock distance is large enough. $r_{\rm in}$ is thus defined by:
\begin{eqnarray}
\int_{r_{\rm in}}^{r_{\rm sh}} {\dd r\over |v|}\equiv{2\pi\over \omega_{\rm f}}.
\end{eqnarray}
A possible strategy to choose the mesh size $\Delta r_{\rm in}$ in the inner region of the flow could be to make sure that the advected perturbations are correctly advected down to this radius $r_{\rm in}$. 
Denoting by $N$ the number of grid zones per wavelength required for an accurate advection and acoustic coupling of vorticity perturbations, the maximal mesh size $\Delta r_{\rm in}$ near the radius $r_{\rm in}$ should be
\begin{eqnarray}
\Delta r_{\rm in}\equiv{1\over N}\;\;{2\pi\over\omega_0}v(r_{\rm in}).\label{critere}
\end{eqnarray}
Our illustration in Fig.~\ref{fig_errP1} suggests $N\sim10$. Of course, the precise value of $N$ depends on the numerical technique used and is expected to vary from code to code but is likely to remain of the same order as our estimate. In any case, Eq.~(\ref{critere}) will be useful for future numerical simulations involving SASI, as a consistency check that the advective-acoustic feedback is properly resolved, at least for the fundamental mode.

\subsection{Mesh size near the stalled shock}

\begin{figure}
\plotone{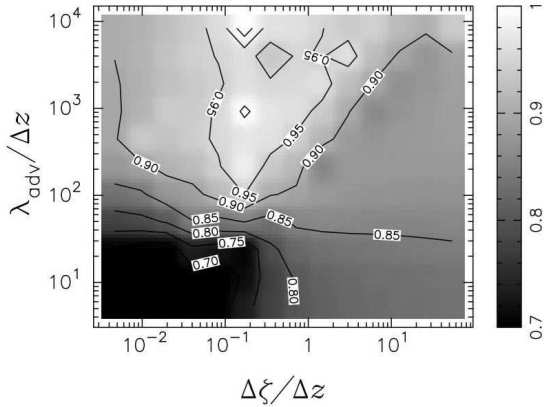}
\caption[dummy]{
Two-dimensional distribution of the power in the a fundamental mode, 
$ |\hat{\delta S_0}/\delta S_{\rm th}|^2$, function of   
$\Delta\zeta/\Delta z$ and $\lambda_{\rm adv}/\Delta z$ obtained in the 1D
simulations of problem 2.}
\label{dzta_lmda}
\end{figure}

Our study of Problem~2 has identified the difficulty 
of accurately calculating the entropy generated by the shock in a
numerical simulations. 
This difficulty is likely to affect any physical quantity depending 
on the physics of the shock, such as the vorticity and the amplitude 
of reflected pressure waves. In this sense, all the numerical simulations 
of core-collapse involving SASI must face a similar difficulty 
with the numerical treatment of the shock. 

We argue that this difficulty is not specific to the linear regime 
of the instability. In the non linear regime of SASI, 
as long as the shock continues to play a fundamental role 
by generating entropy and vorticity perturbations, 
the accuracy of the quantities depending on its behaviour are likely 
to be affected by this first order convergence. However, the details and
precise consequences of this issue in that case remain an open
issue at the present time. Answering these questions will require more
realistic simulations, coupling both problems and carried to the non
linear regime. 

Should the grid size be able to resolve the displacement of the shock for a better accuracy ?
According to the perturbative analysis, the shock displacement $\Delta
\zeta$ is related to the entropy perturbation $\delta S$ by Eq.~(16)
of paper I:
\begin{equation}
\Delta \zeta=\left|
\frac{c_{\rm in}^{2}}{\omega_{0}v_{1}}\;\;\frac{\delta S}{\gamma}
\;\;\frac{1}{\left(1-v_{\rm in}/v_{1}\right)^{2}}
\right|. 
\end{equation} 
We show on Fig.~\ref{dzta_lmda} the accuracy of the numerical simulation, 
compared to the linear calculation, 
depending on how the grid sizes compares to 
both the shock displacement $\Delta\zeta$ and the advection wavelength 
$\lambda_{\rm adv}=2\pi\left|v_{\rm in}\right|/\omega_{0}$, 
in 1D calculations. 
Non linear effects become dominant for $\Delta\zeta>\lambda_{\rm adv}/100$. 
In the linear regime ($\Delta\zeta<\lambda_{\rm adv}/100$), 
an accuracy of $10\%$ requires $\Delta z <\lambda_{\rm adv}/100$. 
Resolving the shock displacement does not seem to be a crucial condition 
for the computation of the entropy production. 

Since the exact properties of numerical convergence vary 
from a numerical scheme to another, 
it is not possible here to determine the real accuracy of existing 
numerical simulations involving SASI. 
At best we can estimate what would be the accuracy of our AUSMDV scheme 
in the conditions used by various authors.
The mesh size $\Delta r_{\rm sh}$ at the radius of the stalled shock 
in published simulations varies depending on their complexity 
and the size of their outer boundary. 
We estimated $\Delta r_{\rm sh}\sim 1$ km in the 2D simulations 
of BM06 and \citet{Scheck+08}, $\Delta r_{\rm sh}\sim 2$ km 
in \citet{Ohnishi+06} and \citet{Iwakami+08}, 
and $\Delta r_{\rm sh}\sim 5$ km in \citet{Burrows+06}.
Estimating the value of the ratio $\lambda_{\rm adv}/\Delta r_{\rm sh}$ 
is possible by identifying $\omega_0$ with the oscillation frequency 
of the dominant mode. 
We estimated $\lambda_{\rm adv}/\Delta r_{\rm sh}\sim 200$ in BM06 
and \citet{Scheck+08}, which seems marginally sufficient to obtain 
a $10 \%$ accuracy from the point of view of Fig.~\ref{dzta_lmda}. 
The discrepancy of $30\%$, noted by FGSJ07 between the numerical results 
of BM06 and the perturbative analysis when the shock distance increases, 
may be related to the fact that the instability becomes dominated 
by the first harmonic rather than the fundamental mode. 
The correspondingly deeper coupling region may require a smaller mesh size, 
as already noted in FGSJ07 on the basis of the structure of the eigenfunction. 
Remembering that the mesh size in BM06 is one of the finest 
among the existing core-collapse simulations, 
particular attention on this issue seems necessary 
for the future simulations in which SASI could play an important role.

\section{Conclusions}
\label{Sec5}

\begin{itemize}
\item A toy model has been used to illustrate through numerical experiments 
the coupling processes described in mathematical terms in paper I. 
Despite the high degree of simplification of our toy model, 
in particular the adiabatic hypothesis and the very local character 
of the deceleration region, these simulations can help us build our intuition 
about the physics of the advective-acoustic instability 
and better recognise it when present in numerical simulations.
\item The results of the perturbative approach have been confirmed 
quantitatively by our numerical simulations. 
\item We have studied the effect of the mesh size on the accuracy 
of the numerical calculation. This will prove useful in the
  future to improve the reliability 
of the hydrodynamical part of simulations involving SASI 
in the core-collapse problem. We have proposed a conservative estimate 
of the desired mesh size close to the neutron star, 
which guarantees that the dominant acoustic feedback 
from advected perturbations is correctly taken into account.
\item The difficulties associated with the numerical
  treatment of the shock have direct consequences on the accuracy with
  which the flow resulting from SASI is calculated: 
without a special numerical effort, the convergence of the computation 
of the growth time and oscillation frequency of SASI is reduced 
to first order even if the numerical scheme converges 
with a higher order away from the shock. 
Among the published simulations of SASI, only the 2D simulations 
with the finest grid seem to be able to estimate the entropy 
and vorticity production at the shock with a $<10\%$ accuracy. 
The importance of an accurate treatment of SASI 
in the core-collapse problem may make it worth implementing 
advanced techniques for the numerical treatment of the shock 
in future simulations, such as the level set method for example
 \citep{SS03}.
 
\end{itemize}

\acknowledgments
The authors are grateful to F. Masset and
M. Liebend\" orfer for their numerical simulations of an early version of
this toy model. Useful discussions with H.-Th. Janka, and constructive 
comments by an anonymous referee are acknowledged.
JS is thankful to R. K\"appeli and N. Ohnishi for helpful comments about the numerical technique. Numerical simulations have been performed with the computational facilities at CEA-Saclay.
This work has been partially funded by the Vortexplosion project
ANR-06-JCJC-0119.



\begin{thebibliography}{}
\bibitem[Blondin et al.(2003)]{Blondin+03}
Blondin, J. M., Mezzacappa, A., 
\& DeMarino, C. 2003, ApJ, 584, 971
\bibitem[Blondin \& Mezzacappa (2006)]{BM06}
Blondin, J. M., \& Mezzacappa, A. 2006, ApJ, 642, 401 (BM06)
\bibitem[Blondin \& Mezzacappa (2007)]{BM07}
--- 2007, Nature, 445, 58
\bibitem[Burrows et al.(2006)]{Burrows+06}
Burrows, A., Livne, E., Dessart, L., Ott, C. D., 
\& Murphy, J. 2006, ApJ, 640, 878
\bibitem[Colella \& Woodward (1984)]{CW84}
Colella, P., \& Woodward, P. R. 1984, J. Comput. Phys., 
54, 174
\bibitem[Foglizzo (2008)]{F08}
Foglizzo, T. 2008, submitted to ApJ (paper I)
\bibitem[Foglizzo et al.(2007)]{Foglizzo+07}
Foglizzo, T., Galletti, P., Scheck, L., \& Janka, H.-Th. 2007, ApJ, 654, 1006 (FGSJ07)
\bibitem[Iwakami et al.(2008)]{Iwakami+08} 
Iwakami, W., Kotake, K., Ohnishi, N., Yamada, S., \& Sawada, K. 2008, ApJ, 678, 1207 
\bibitem[Laming (2007)]{Laming07}
Laming, J. M. 2007, ApJ, 659, 1449
\bibitem[Laming (2008)]{Laming08}
Laming, J. M. 2008, Erratum to be published in ApJ
\bibitem[Liou \& Steffen (1993)]{LS93}
Liou, M. -S., \& Steffen, C. J. 1993, J. Comput. Phys., 
107, 23
\bibitem[Marek \& Janka (2007)]{MJ07}
Marek, A., \& Janka, H.-Th. 2007, ApJ, submitted (arXiv: 0708.3372)
\bibitem[Miyoshi \& Kusano (2005)]{MK05}
Miyoshi, T., \& Kusano, K. 2005, JCoPh, 208, 315 
\bibitem[Murphy \& Burrows (2008)]{MB07}
Murphy, J. W., \& Burrows, A. 2008, ApJ, 688, 1159
\bibitem[Ohnishi et al.(2006)]{Ohnishi+06}
Ohnishi, N., Kotake, K., \& Yamada, S. 2006, ApJ, 641, 1018
\bibitem[Scheck et al.(2008)]{Scheck+08}
Scheck, L., Janka, H.-Th., Foglizzo, T., \& Kifonidis, K. 
2008, A \& A, 477, 931 
\bibitem[Scheck et al.(2004)]{Scheck+04}
Scheck, L., Plewa, T., Janka, H.-Th., Kifonidis, K., \& M\"{u}ller, E. 
2004, Phys. Rev. Lett., 92, 011103 
\bibitem[Sethian \& Smereka (2003)]{SS03}
Sethian, J.A., \& Smereka, P., 2003, Annual Review of Fluid Mechanics, 35, 341
\bibitem[Teyssier (2002)]{T02}
Teyssier, R., 2002, A\&A, 385, 337
\bibitem[Toro et al.(1994)]{T+94}
Toro, E. F., Spruce, M., \& Speares, W., 1994, Shock Waves, 4, 25
\bibitem[Toro (1997)]{toro97}
Toro, E. F., 1997, Riemann solvers and numerical methods for fluid
dynamics (Springer)
\bibitem[Wada \& Liou (1994)]{WL94} 
Wada, Y., \& Liou, M. S. 1994, AIAA Paper, 94-0083
\bibitem[Yamasaki \& Foglizzo (2008)]{YF08} 
Yamasaki, T., \& Foglizzo, T. 2008, ApJ, 679, 607
\bibitem[Yoshida et al. (2007)]{Y07} 
Yoshida, S., Ohnishi, N., \& Yamada, S. 2007, ApJ, 665, 1268
\end{thebibliography}
\end{document}